# Developing high energy mode-locked fiber laser at 2 micron

C. Huang[1], C. Wang[1], W. Shang[1], Y. Tang[1,2*], and J. Xu[1,2]

[1]Key Laboratory for Laser Plasmas (Ministry of Education) and Department of Physics and Astronomy, Shanghai Jiao Tong University, Shanghai 200240, China

[2]IFSA Collaborative Innovation Center, Shanghai Jiao Tong University, Shanghai 200240, China

Correspondence and requests for materials should be addressed to Y.T. (yulong@sjtu.edu.cn)

**Abstract**

**While dissipative soliton operation has successfully improved the pulse energy of 1 μm and 1.5 μm fiber lasers to tens of nanojoules, it is still hard to scale the pulse energy of dissipative solitons at 2 μm due to the anomalous dispersion of the gain fiber. Based on theoretical simulation, we analyze intracavity dynamics of dissipative solitons (DSs) and propose that gain fiber should be condensed to short length in order to scale the pulse energy of 2 μm DSs. The simulation predicts pulse energy of over 10 nJ for 2 μm dissipative solitons, comparable to that achieved in the 1 μm and 1.5 μm regimes. Experimental operation generates stable 2 μm DSs from a linear cavity with pulse energy of 4.9 nJ and dechirped pulse duration of 579 fs. These results advance our understanding of mode-locked fiber laser at different wavelengths and lay an important step in achieving high energy ultrafast laser pulses from anomalous dispersion gain media at 2 μm.**

Realizing high energy ultrashort pulses in various wavelength regions is very important for a variety of scientific and industrial applications and has been the persistent pursuit of scientists. Therefore, different mode-locking mechanisms have been extensively explored to deliver high energy ultrashort pulses[1-5]. Among these, the dissipative solitons (DS) mode-locking takes advantage of balance between not only nonlinearity and dispersion but also gain and loss, giving rise to pulse energies 1~2 orders of magnitude larger than that from conventional soliton mode-locking[1]. Up to now, the DS pulse energies at 1 μm and 1.5 μm from fiber systems have exceeded 20 nJ with femtosecond pulse duration[6-8].

Recent years, ultrashort pulses at 2 μm have attracted considerable interest and have played important roles in extensive areas like LIDAR, surgical operation, molecule spectroscopy, remote sensing, etc.[9]. Contrary to the encouraging results of DSs in the 1 µm and 1.5 µm regions, the pulse energies of 2 µm fiber lasers still remain at a low level. This is because the readily available gain fibers in the 2 µm region show relatively large anomalous dispersion, which makes most of these lasers operating in the conventional soliton regime. By inserting elements (such as a grating) with small normal dispersion into the cavity, pulse energy can be improved to certain extent, but the underlying mechanism makes this kind of fiber laser always limited by the soliton area theorem[10].

For the purpose of breaking through the pulse energy bottleneck, some researchers have tried to employ normal dispersion fiber components to construct DS fiber lasers at 2 µm. With a chirped fiber Bragg grating providing normal dispersion, Gumenyuk et al. demonstrated 2 µm DSs in a semiconductor saturable absorber mirror (SESAM) mode-locked thulium/holmium fiber laser with pulse energy of 2.2 nJ[11]. However, its relative narrow spectral bandwidth (5 nm) limits the shortest dechirped pulse duration to ~1 ps. It is well known that the dispersion of a fiber can be shifted through adjusting fiber core diameter and numerical aperture (NA)[12], thus specially designed fibers can provide normal dispersion at the 2 µm region. Using a very small core (2.75 µm) and high NA (0.28) dispersion compensating fiber (DCF), Haxsen et al. reported a SESAM/nonlinear polarization evolution (NPE) hybrid mode-locked 2-µm fiber laser[13]. Although the output pulse can be dechirped to 482 fs, the pulse energy is only 0.67 nJ, comparable to conventional solitons. Q. Q. Wang et al. also reported a 2 µm DS fiber laser with a similar DCF to manage intracavity dispersion and obtained 0.45 nJ, 2.3 ps pulses[14].

Here, we present a detailed investigation into the intracavity balance and dynamics of the 2 µm normal-dispersion fiber laser, uncovering an important difference between 2 μm normal dispersion fiber lasers and their 1 μm and 1.5 μm counterparts. Because of this difference, DSs at 2 μm, though based on the same theory, have been limited to low energy so far. According to theoretical analysis, we propose that the anomalous gain fiber should be condensed as short as possible to efficiently decouple gain from dispersion and nonlinearity. In this way, the pulse can avoid too much phase accumulation and thereby evolve stably into high energy DS. Numerical simulations show that over 10 nJ 2 μm DSs can be realized within a linear cavity. To verify this, we build a SESAM mode-locked thulium-doped fiber laser and obtain 4.9 nJ DS pulses with 579 fs dechirped pulse duration. To the best of our knowledge, this is the highest energy DSs with femtoseconds from 2 μm single mode fiber oscillator.

A schematic model for the mode-locked laser is illustrated in Fig. 1a. The dispersion compensating fiber (DCF), single mode fiber (SMF), and gain fiber (GF) complete the dispersion map. Numerical simulations are based on the well-known nonlinear Schrodinger equation (NLSE) with gain[1-2]. The SMF-28 (8.2/125 μm, 0.14 NA) has a length of 1.4 m, with $β_2$ = -67 ps$^2$/km and γ = 0.001 (Wm)$^{-1}$, while the DCF (2.2/125 μm, 0.35 NA) is 1.5 m long with $β_2$ = 93 ps$^2$/km and γ = 0.007 (Wm)$^{-1}$, respectively. The 0.2 m GF (5/125 μm, 0.24 NA), with $β_2$ = -12 ps$^2$/km and γ = 0.003 (Wm)$^{-1}$, has the gain g = $g_0$ / [1 + $E_{pulse}$/$E_{sat}$ + (ω-$ω_0$)$^2$/Δω$^2$], where $g_0$ corresponds to ~30 dB of small-signal gain, $E_{sat}$ is the gain saturation energy, and the gain bandwidth is assumed to be ~90 nm. The saturable absorber (SA) is simulated by the transfer function T = 1 - $l_0$/[1 + P(τ)/$P_{sat}$], where the unsaturated loss $l_0$ is set to 0.7, P(τ) is the instantaneous pulse power, and $P_{sat}$ is the saturation power. A spectral filter (SF) with 150 nm bandwidth is inserted to describe the bandwidth of SA. The equations are solved with the split-step Fourier method. Start with white noise, the calculation proceeds until a steady state is reached. A typical solution with pulse energy of 5 nJ is obtained as shown in Fig. 2.

Owing to the joint effects of normal group velocity dispersion (GVD) and nonlinearity in DCF, the pulse propagates with its duration increases monotonically. The weak broadening in the temporal profile is compressed by SMF and GF segments with anomalous GVD. The spectral bandwidth of the pulse with steep edges is nearly static in the cavity. However, its top exhibits sharp peaks near edges after the pulse amplifies in the GF and then undergoes self-phase modulation in the following DCF. Then the spectrum is reshaped by SA, DCF, GF, and SMF, successively, and evolves back to near-flat top. Fig.2 (b) shows the temporal phase evolution in the three kinds of fibers (DCF, GF and SMF). We see that there is only a small amount of phase shift induced by the short GF, which is effectively compensated by the other two kinds of fibers.

To gain deeper insight into the intracavity pulsing dynamics, a qualitative illustration for 2 μm DSs is summarized in Fig.3, along with their 1 μm and 1.5 μm counterparts (inset)[5]. With normal dispersion, the GFs (e.g. Yb-doped and Er-doped fibers) introduce positive phase shift, which can be compensated by the negative phase shift provided by the SMF, even if the shift is relatively large (see the inset of Fig.3). However, the scenario is quite different for the 2 μm wavelength regime. The GFs at 2 μm provide anomalous dispersion and thus negative phase shift. To obtain ultrafast DSs at 2 μm, DCF and SMF are both necessary for the cavity, and the net dispersion should not be too large. For a given cavity, the tolerance of phase shift left for GF is limited to a relatively small value (purple area). By increasing fiber's length, the phase shift incurred by DCF or SMF is small (yellow or green dashed arrow) while phase shift induced by GF is significant (red dashed arrow). A short GF has larger slope in the illustration and thus can achieve higher energy within the phase limitation region (red arrow). On the contrary, due to smaller slope, a long GF has to sacrifice a large part of amplitude (thus energy) to reduce phase shift under the tolerable level (orange arrow). Therefore, stable pulses from a cavity with a longer GF have lower energies.

In the light of the above analysis, we propose using a condensed gain-fiber (shortened to a small length while at the same time provides high adequate gain) for scaling DSs at 2 μm and mid-infrared spectral regions. Under this condition, the gain fiber will have a large slope (Fig. 3), and thus can provide high enough pulse energy, keeping the phase accumulation within the limitation range at the same time.

To confirm the above idea, simulations are also performed in the above mode-locked fiber laser with various GF lengths. The simulated maximum output pulse energies with various GF lengths are indicated in Fig.4. It is clear that decreasing the GF length will dramatically increase the pulse energy. For instance, as high as 11 nJ pulse is supported by the cavity with a 0.2 m GF and can be dechirped to ~300 fs outside the cavity. This confirms our expectation for achieving high energy ultrashort pulses with anomalous dispersion gain fibers: shortening the anomalous GF to a sufficiently small length can significantly improve the pulse energy. In addition, this short GF proposal is also valid for ring cavities.

Encouraged by the simulation results, a net normal dispersion fiber laser at 2 μm is constructed to realize high energy DSs. The experimental setup for the mode-locked fiber oscillator with a linear cavity is schematically shown in Fig.1 (b). The pump source is a CW erbium/ytterbium-codoped fiber laser (EYFL) with maximum output of 1 W centered at 1550 nm. The pump light is delivered into the gain fiber by a 1550/2000 nm wavelength division multiplexing couplers (WDM), which is made by SMF-28 fiber. The coupling efficiency is ~95% with splice loss included. Considering the trade-off between enough gain and less phase shift, 15 cm single-cladding Tm-doped fiber is chosen to decouple gain from GVD and nonlinearity. The dispersion of these three kinds of fibers is consistent with the parameters used in the simulation. Consequently, the total net normal dispersion of the cavity is ~0.04 $ps^2$. On the output end, the perpendicularly cleaved fiber facet

(~4% Fresnel reflection) is employed to provide laser feedback and acts as the output coupler. In this way, most of the intracavity energy can be extracted. The other end of the cavity fiber is directly butt coupled to the SESAM with central wavelength of 2000 nm, modulation depth of 25%, and saturation fluence of 35 mJ/cm$^2$. The simple design of the cavity offers great flexibility for future modifications or integration into other function systems. The laser reached its continuous wave (CW) threshold with pump power of ~500 mW. When pump power is increased to ~650 mW, stable CW mode locking is self-started. Remarkably, no Q-switching or Q-switched mode-locking regime is observed. This is due to the large output coupling ratio, which suppresses the intermediate transitions between the CW laser operation and the CW mode locking regime[15]. With the maximum pump power of 1 W, the maximum average output power of 158 mW for the 2 μm DS is obtained. The Fig.5 (a) displays the oscilloscope trace (measured with a 12.5 GHz detector and a 2.5 GHz oscillator) of the mode-locked pulse trains at the maximum output level in 50 ns/div time scale, showing a repetition rate of ~32 MHz (corresponding to the round-trip time of ~31 ns). Consequently, the maximum single pulse energy is ~4.9 nJ. The typical laser spectrum (at the highest output value) was analyzed by a mid-infrared spectrum analyzer with a spectral resolution of 0.1 nm, as shown in Fig.5 (b). The wavelength is centered at 1918 nm with 3 dB spectral width of 15 nm. Steep spectral edges indicate the typical characteristics of DSs[1]. The radio-frequency (RF) spectrum is about ~52 dBm signal-to-noise ratio (Fig.5 (c)). Fig.5 (d) shows the autocorrelation trace of the pulses directly produced from the cavity at maximum output. Assuming a Gaussian shape, it corresponds to pulse duration of 16 ps. The time-bandwidth product is calculated as 19, which is a large departure from the Fourier transform limited value. The highly linearly chirped pulse is dechirped to 579 fs by a SMF-28 fiber (see Fig.5 (e)), and then the time-bandwidth product is reduced to 0.7.

According to simulations results, to achieve high energy DSs, the GF length should be as short as possible to suppress the phase shift while at the same time provides enough gain. To this end, highly doped GF (short length with enough gain) should be adopted to complete the cavity. Such kind of GF can efficiently decouple gain from dispersion and make phase shift in the GF negligible. This can efficiently avoid the process of pulses evolution to conventional solitons during amplification in the GF. Guided by the theoretical model, we experimentally demonstrate a net normal dispersion mode-locked Tm-doped fiber laser. The laser delivers 4.9 nJ DSs with pulse duration of 579 fs after dechirped. To the best of our knowledge, this is the highest energy for 2 µm femtosecond pulses directly generated from a fiber oscillator, and more than twice the pulse energy of the previous results[11]. Although the pulse energy from this initial experiment is high for 2 µm DSs, it is still much lower than that from the 1 µm and 1.5 µm counterparts. In the current experiment, the output pulse energy is limited by the available pump power. For further energy scaling of 2 µm DSs, 1.5 µm laser should be amplified to a high power level with an Er/Yb fiber amplifier. Due to the lower cost, smaller size, and higher power, laser diode at 793 nm is also a good option. On the other hand, more condensed GFs should be adopted for the linear cavity. The recently reported highly Tm-doped silicate glass fibers[9] (a gain per unit length of greater than 2 dB/cm) make the route feasible. Therefore, ultrafast pulse at 2 µm with energy exceeding 20 nJ is very probable in the near future.

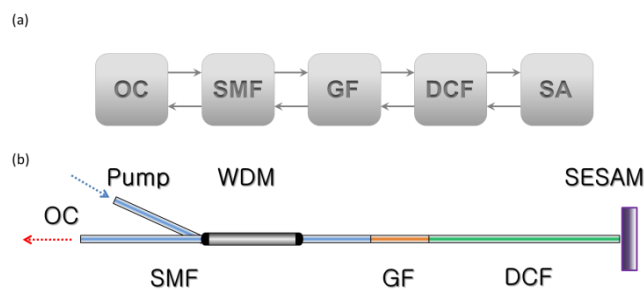

Figure 1 (a) Schematic of a 2 μm normal dispersion fiber laser with a linear cavity. DCF, dispersion compensating fiber; GF, gain fiber; SMF, single mode fiber; OC, output coupler; SA, saturable absorber. (b) Experimental setup of mode-locked Tm-doped fiber laser.

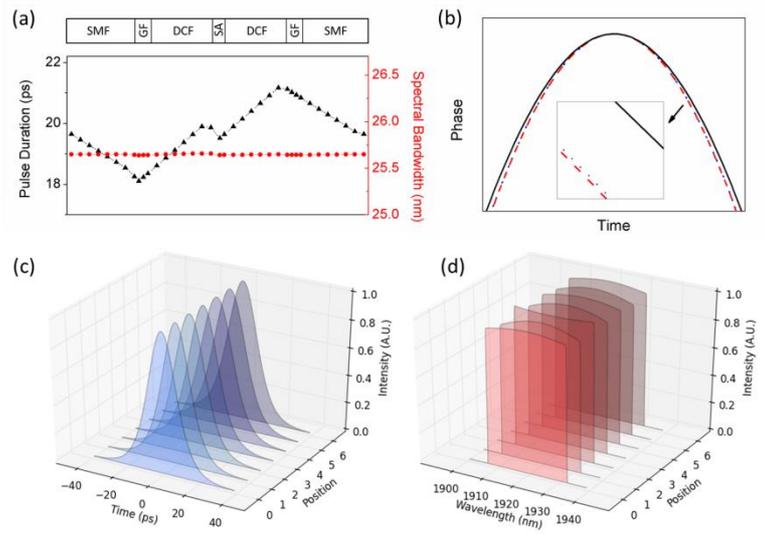

Figure 2 Intracavity Dynamics. (a) Pulse duration (black triangles) and spectral bandwidth (red circles) evolution. (b) Temporal phase of the solution to the laser plotted after DCF (black solid), after GF (red dashed), after SMF (blue dotted). (c) Temporal and (d) spectral profiles of the pulse after SMF, GF, DCF, SA, DCF, GF, SMF, respectively.

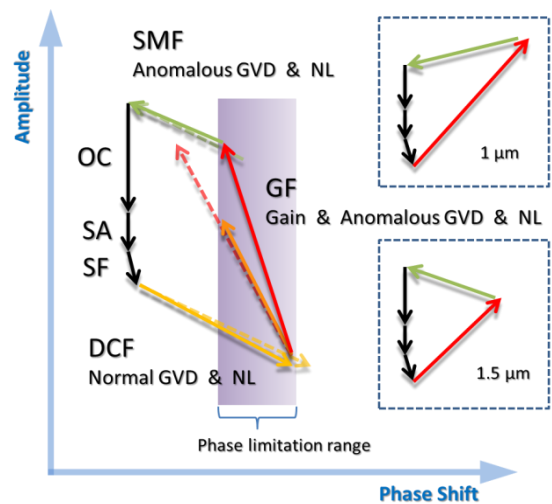

Figure 3 Qualitative illustration of the amplitude and phase balances in a DS fiber laser at 1 μm, 1.5 μm (insets), and 2 μm

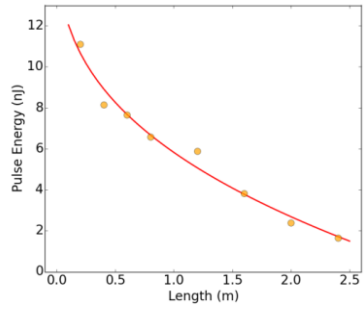

Figure 4 Calculated maximum pulse energy versus the length of GF.

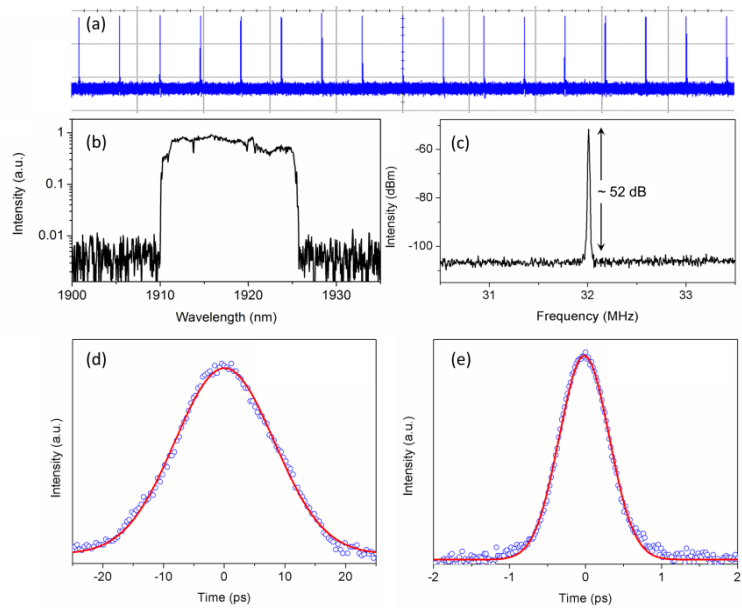

Figure 5 (a) Pulse train of the mode-locked laser, 50 ns/div. (b) Laser spectrum and (c) RF spectrum of mode-locked Tm-doped fiber laser. Autocorrelation of the pulse (d) before and (e) after compression.